\documentclass[12pt]{scrartcl}
\usepackage[]{graphicx}
\usepackage{psfrag}
\usepackage{amsfonts}
\usepackage{xspace}

\usepackage[english]{babel}
\usepackage{url}

\usepackage{listings}
\usepackage{cite}
\newcommand{\be}{\begin{eqnarray*}}
\newcommand{\ee}{\end{eqnarray*}}
\newcommand{\ben}{\begin{eqnarray}}
\newcommand{\een}{\end{eqnarray}}

\newcounter{bla}

\newcommand{\fmdf}{Full-Metadata Format\xspace}
\newcommand{\fmf}{FMF\xspace}

\begin{document}
  \makeatletter
  \newcommand*{\footref}[1]{
  \begingroup
    \unrestored@protected@xdef\@thefnmark{\ref{#1}}%
    \endgroup
   \@footnotemark}
  \makeatother

\title{On the Communication of Scientific Results: The Full-Metadata Format}
\author{Moritz Riede\thanks{Institut f\"{u}r Angewandte Photophysik, Technische
  Universit\"{a}t Dresden, George-B\"{a}hr-Str. 1, 01069 Dresden,
  Germany \protect\label{IAP}}
 \and Rico Schueppel\footref{IAP}
 \and Kristian O. Sylvester-Hvid\thanks{Ris\o{} National Laboratory, Technical
    University of 
  Denmark, Frederiksborgvej 399, 4000 Roskilde, Denmark}
 \and Martin K\"{u}hne\thanks{Freiburger Materialforschungszentrum, Universit\"{a}t
  Freiburg, Stefan-Meier-Str. 21, 79104 Freiburg, Germany
  \protect\label{FMF}} 
\and Michael C. R\"{o}ttger\footref{FMF} 
\and Klaus Zimmermann\footref{FMF}
\and Andreas W. Liehr\thanks{coresponding
  author: liehr@users.sourceforge.net}\footref{FMF}}

\date{\today}
\maketitle
\begin{abstract}
In this paper, we introduce a scientific format for text-based
data files, which 
facilitates storing and communicating tabular data sets. The so-called
Full-Metadata Format builds on the widely used INI-standard and is based on four
principles: readable self-documentation, flexible structure, fail-safe
compatibility, and searchability. As a consequence, all metadata required to interpret the tabular
data are stored in the same file, allowing for the
automated generation of publication-ready tables and graphs and the
semantic searchability of data file collections. The
Full-Metadata Format is introduced on the basis of three comprehensive examples. The
complete format and syntax is given
in the appendix.  
\end{abstract}

\maketitle

\section{Introduction}
\label{sec:introduction}

In the last few years an increasingly sophisticated experimental infrastructure has
evolved enabling scientists to share not only knowledge but also primary
data via scientific
publications \cite{Klump2006_dataPublication,Uhlir2007_openData,ESSD}. With this
increase in sharing primary or processed scientific data the
lack of intuitive and well defined data formats for simple tabular
data has become increasingly obvious.
For complex data sets like the ones dealt with in the
earth sciences, adequate binary formats like the Network Common Data Form
(netCDF~\cite{netcdf}) or the Hierarchical Data Format
(HDF~\cite{hdf5}) are well established \cite{netcdf_usage,hdf5_usage}, and the publication of
observational geophysical data in World Data Centres has developed into an
effective mechanism for the exchange of data
\cite{NationalResearchCouncil1997_OpenData}. Another example is the
information technology infrastructure for handling the data of the
ATLAS experiment \cite{Nowak2008_ATLAS}, where the event data is
mainly stored in the ROOT file format \cite{Root2008_5.21}. For 
less complex data structures, like tabular data as typically
encountered in many parts of natural and technical sciences, no single standard format has evolved.
 
The success of the HDF and netCDF relies on the fact that
the formats are well defined and integrate smoothly into the workflow
of scientists in different laboratories. Although these formats
are capable of storing and documenting simple tabular data, the 
overhead of work needed to process binary files generally poses a barrier to the
use of these formats in fields where complex data structures are
seldom dealt with.

A natural requirement of a standardized file format for tabular data is that it
allows scientists to add observations, notes, parameter
specifications and analysis results by editing 
in clear text using any given text editor. This constitutes what most
of the
overwhelming number of data formats used in laboratories around the
world have in common. However, as text files are easy to handle,
every laboratory, working group or even scientist has an individual
standard of documenting scientific results with text-based
formats. While this is completely sufficient in a short term
perspective, it becomes intractable with the tendency of research
projects to rely on the cooperation of international
consortiums involving many different laboratories. Furthermore,
in publishing scientific results, there is an increasing demand to
provide also processed data as supplementary data or to even publish
primary data in OpenData
repositories~\cite{Uhlir2007_openData}. Thus, there is a need
for a common data format for tabular data which is: 
\begin{description}
\item[Readable and self-documenting:]
The data should be written in the same way the scientist is used to
reading it, as e.g. in a laboratory notebook. It should be clear, text
based and processable with any word processing tool. The file
format should include sections which allow the scientist to document
the data and its origin, and this to such an extent that no other source be required to
to understand the origin of the data. This standard also implies that the data files are search-able and
individual data sets can be tracked down by semantic or keyword based queries. 
\item[Flexible but structured:]
The data format must be flexible enough to allow the individual
scientist to structure and classify data in an intuitive and
convenient way without compromising the overall structure and
readability as stipulated by the format. The overall structure of the
format must be such that data files may still be processed with common
analysis and visualisation software packages, thus facilitating the
automated processing of data from different measurement sources and
measurement series. This further implies that format and syntax
specifications are largely decoupled such that annotated data may
smoothly cross language zones. 
\item[Fail-safe and compatible:]
A fail-safe data format has to assure that the format is robust against
misinterpretation by a parser or deviations from the format specifications. As the
format specification is expected to evolve in time, backwards
compatibility must always be retained. 
\item[Searchable:] Communicating scientific results implies that relevant data sets
can be found within a certain collection by means of
simple queries. This requires the documentation of scientific data in the form of self-documenting file formats.
Further, a collection of scientific data files must be catalogued not only according to bibliographic items or keywords,
but also to physical quantities. 
\end{description}
It is of paramount importance that the data format integrates smoothly
into the existing workflow of the
scientist and supports the natural working cycle of collecting and
structuring information. To become widely accepted, the threshold of
annotating primary data with additional information must be as low as
possible; scientists should not have to start learning a complex
syntax or a sophisticated mark-up language, which for all
practical purposes will require specialised software tools.

In the following we present a syntax for a self-documented scientific data
file format for 
tabular data sets which we call the \textit{\fmdf} (\fmf). It is 
purely text based and \fmf-files consist of two parts:
the first part contains  
the metadata describing the data written in the second
part of the file. Because most scientific software tools 
support the skipping of some initial header lines, the data stored in
the \fmf-file can directly be processed as
usual. Yet, the documentation of the data remains always at hand. The
proposed file format has evolved from the development of 
high-throughput experimental setups for the processing and
characterisation of organic solar cells
\cite{SPIE2006,phdriede,Riede2008_OSC}, and is applicable to all kinds
of tabular data encountered in the natural sciences and engineering. A
further demonstration of the capabilities of the \fmf-format is constituted by its
incorporation into the scientific analysis software
Pyphant \cite{Pyphant,Zimmermann2008_grammar}, which supports the
computation with units 
and the analysis of metadata
\cite{Hanko2006_Pyphant,Bruns2008_Pyphant}.

Below the \fmdf{} is first described by means of two examples highlighting
its principles and potential. A third example sketches the
capabilities of searching the metadata of \fmf-files for relevant data
sets. In the appendix
the complete format and syntax definitions are listed.

\section{A Basic Example: Communicating Simple Tabular Data}
\label{sec:case_study_I}
In this example a typical data exchange between
two work groups is considered to demonstrate the benefit of human readable data
formats for the communication between scientists.
The goal of the cooperation might be the
enhancement of the power conversion efficiency of organic solar cells,
or the numerical modelling of the 
characteristics of solar cells with respect to production parameters.
This requires exchanging data between the groups. 
In Fig.~\ref{fig:bare}a the screenshot of a typical data set of a current-voltage characteristic is shown, which
is formatted in the most common data format for tabular data: pure
columns of numbers. The corresponding graphics is shown in Fig.~\ref{fig:bare}b.
The missing axis labels
indicate that important information like the name, symbol, and units
of the plotted physical quantities are not provided with the data set,
thus only allowing for a qualitative assessment of the data.
\newlength{\Bildhoehe}
\settoheight{\Bildhoehe}{\includegraphics[width=0.382\textwidth]{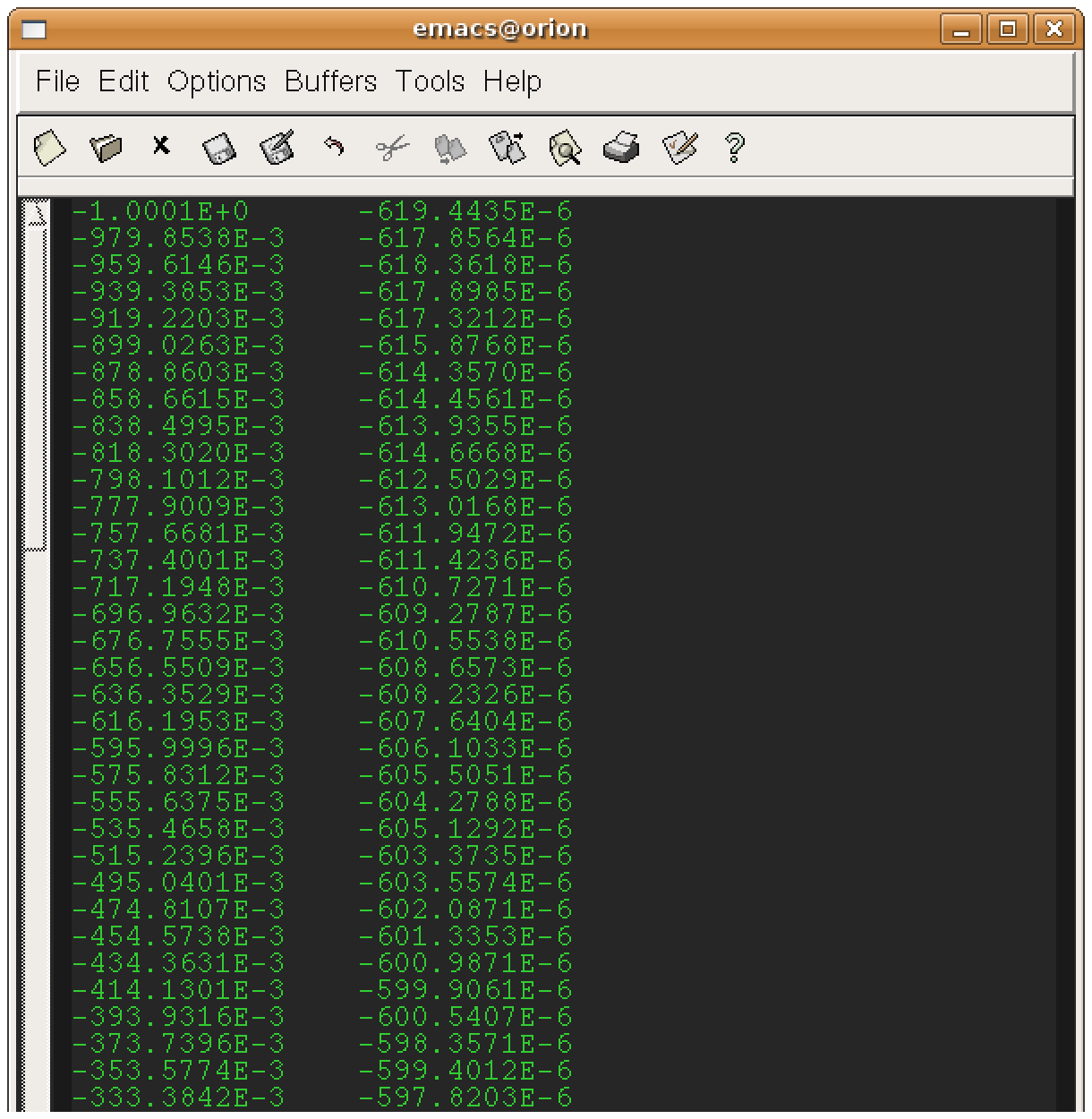}}
\begin{figure}
  \begin{center}
  \begin{tabular}{ll}
    (a) & (b)\\
  \includegraphics[width=0.382\textwidth]{Riede2008_20060417_0129-bare-screenshot}
&
  \includegraphics[height=\Bildhoehe]{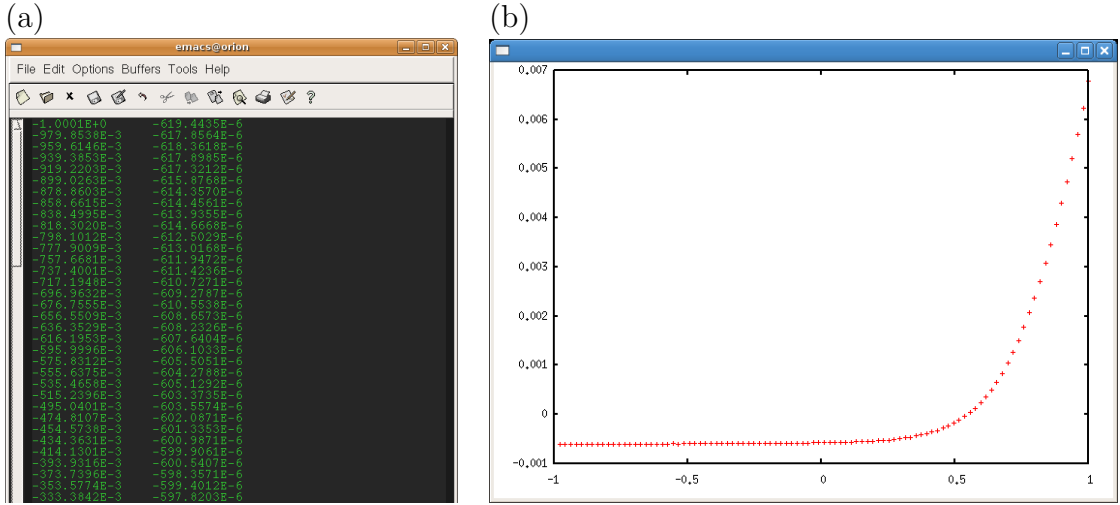}
\\
 \end{tabular}
  \end{center}
\caption{\label{fig:bare}A typical data file exchanged between
  scientists. (a) Screenshot of a text editor's view of the data
  file and (b) the corresponding plot with a qualitative relation from
  the values listed in the data file.
}
\end{figure}

\begin{figure}
  \begin{center}  
  \includegraphics[width=0.618\textwidth]{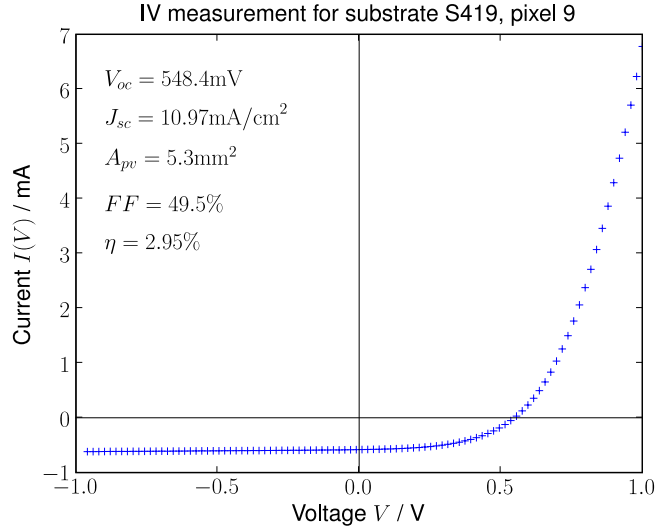}
  \end{center}
\caption{\label{fig:poor} Publication-ready graphics of an
  IV-characteristic based on a \fmdf{} file (Fig.~\ref{fig:FMF_solar_example}): IV measurement for substrate
   S419, pixel 9. The solar cell characteristics are measured under illumination with a mismatch corrected intensity of $I_{AM1.5} = 100 \mathrm{mW/cm}^2$.
}
\end{figure}

This lack of information can be clarified by a phone call or an email.
Typically, the response to such requests depends on how the working group
internally documents the primary data. It may either be well
documented by means of protocols in the laboratory notebook of the
scientist in charge, but the protocol has not been attached to the
e-mail, or the data file format is standardised by an internal format
convention of the working group, but the format has not been
documented. In both cases a useful response would at least communicate
that the 
first column is voltage $V$ in units of Volt, the second column is current $I$ 
  in units of Ampere, the current $I$ is measured as a function of $V$, the
  device has an active area $A_{pv}$ of 5.3mm$^2$, and is exposed to an
  illumination intensity $I_{AM1.5}$ of
  100mW/cm$^2$~\cite{OSC_mismatch}. Having this information at hand,
  the diagram can be labelled 
correctly as required for further processing, publication, and understanding (Fig.~\ref{fig:poor}).
In addition, characteristic properties like the fill factor
($FF=45.5\%$) and the power conversion efficiency ($\eta=2.95\%$) can
be extracted from the data \cite{OSC_mismatch}. However, this is only a temporary solution, as the original data file
is unlikely to be annotated accordingly. The next time the data set is
used, the same questions will arise. The data set might even become completely
useless, if the relevant protocol of the laboratory notebook cannot be
identified anymore or if the person responsible for the measurement cannot clarify the units~\cite{Kuehne2008_knowledge}.

\begin{figure}
  \centering
  \psfrag{qq}{$\vdots$}
  \includegraphics[width=0.618\textwidth]{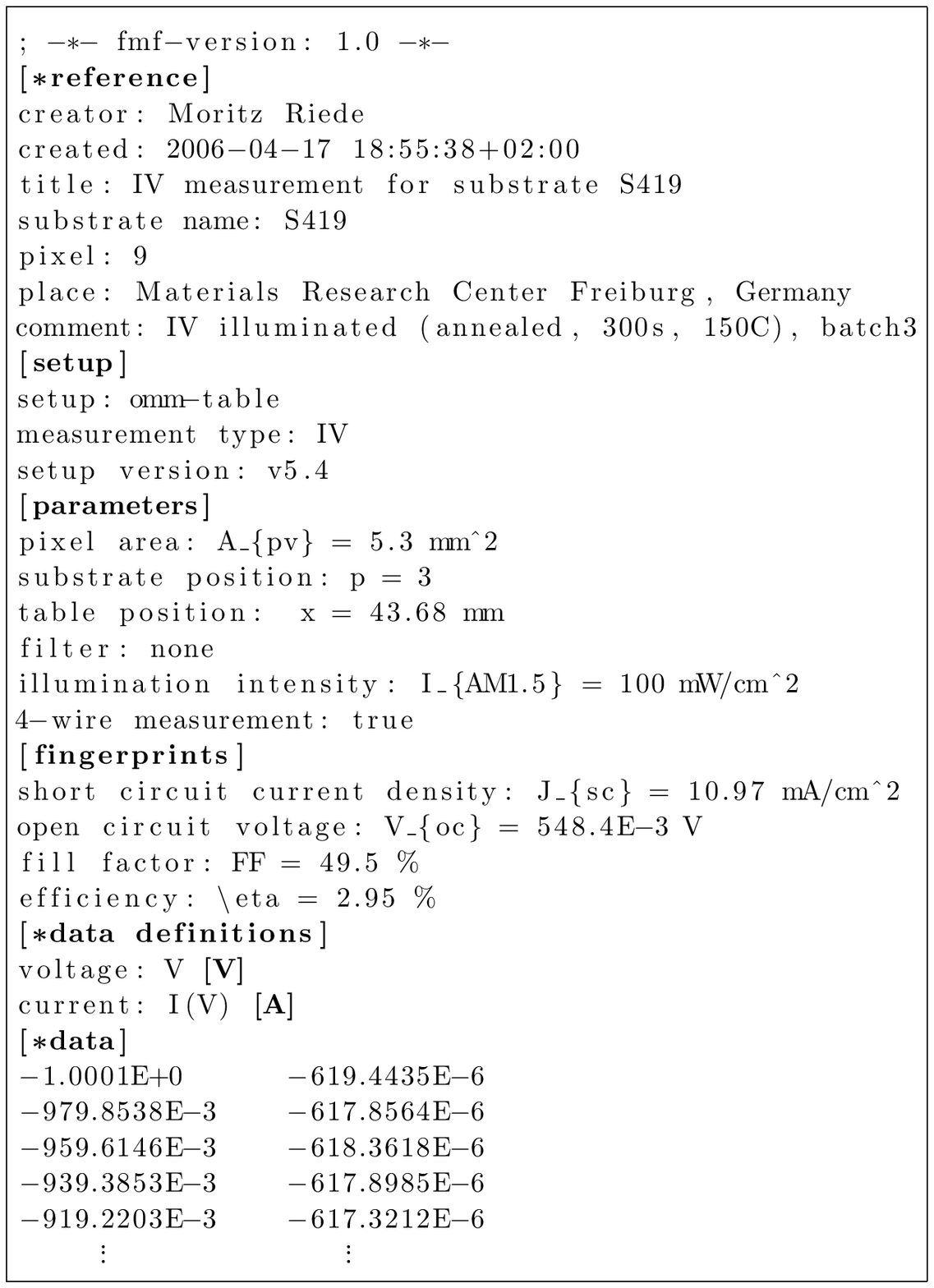}
\caption{\label{fig:FMF_solar_example} The first lines of a
  self-documented data file in the \fmdf{} \cite{Riede2009_SI20090303a}, cut after some tabular data values.
}
\end{figure}

Clearly, it would have been better to annotate
the data set with the missing information right from the
start. Using the proposed format, the \fmf-file corresponding to the
data depicted in Fig.~\ref{fig:poor} is shown
in Fig.~\ref{fig:FMF_solar_example}. Some metadata and the first few
lines of the raw data are shown. The list of metadata is not exhaustive and only as
much is shown as to highlight the possibilities of the file
format. Note, that the proposed file format has similarities with the
INI file format, but goes beyond this in its possibilities due to extra
rules. The detailed syntax is given in~\ref{sec:syntax}. 

The file shown in Fig.~\ref{fig:FMF_solar_example} starts with a
single line describing
the version of the \fmdf{}.
The next part contains all the metadata required for understanding
the actual data. 
This metadata is given in a simple and user-friendly way by
structuring the file into sections.  
Bibliographic information resides in section \lstinline![*reference]!,
column definitions in 
section \lstinline![*data definitions]!, and the corresponding columns of data in section
\lstinline![*data]!. The bibliographic information is either used
for internal archiving purposes or for publishing the data file in an
OpenData repository~\cite{Uhlir2007_openData} like for example \cite{Riede2009_SI20090303a}. These three sections
are mandatory and comprise the fundamental structure of a
\fmdf-file. 

The other sections in the example in Fig.~\ref{fig:FMF_solar_example}, \lstinline![setup]!,
\lstinline![parameters]!, and \lstinline![fingerprints]! are not preceded by an asterisk. These are user defined sections and can contain arbitrary extra metadata.
All sections, except the \lstinline![*data]! section, contain items 
coded as colon separated 
\begin{equation}
  \label{eq:key-value}
  key: value
\end{equation}
pairs. The $key$ cannot be a colon, because the first colon per line separates $key$ and
$value$.
A $value$ can be boolean, numerical, a quantity, a timestamp, or a
string. In \lstinline![*data definitions]! section the value must
be a column specificator (\ref{sec:tables}).

According to the meta data, the file shown in Fig.~\ref{fig:FMF_solar_example} was created by \textit{Moritz Riede} on
\textit{17th of April 2006} at \textit{18:55:38} local time, which is 2
hours ahead of UTC (cf. Tab.~\ref{tab:DateTime}).
It contains data for the solar cell on pixel
\textit{9}, located on a substrate with the unique identifier
\textit{S419}. This identifier
can be used for referencing the processing and measurement history of
the solar cell \cite{phdriede,Kuehne2008_knowledge}. A short comment completes the \lstinline![*reference]! section.

The section \lstinline![setup]! is used in the example to describe the
measurement type and the setup used. Many measurements can be carried out
on different setups, each with their own distinct features, which are relevant when interpreting the data~\cite{phdriede, Riede2008_OSC}.
A set of important measurement parameters important to the interpretation of the data are recorded within the
section \lstinline![parameters]!. A special mention should be given to key-value pairs which we characterize as \textit{quantities}, and in section \lstinline![parameters]! for example, the active area of the solar cell is specified as: 
\begin{verbatim}
pixel area: A_{pv} = 5.3 mm^2
\end{verbatim}
It is written like a typical parameter specification and comprises a
name (``pixel area''), a symbol in \LaTeX{}-notation ($A_{pv}$) \cite{Lamport1994_latex}, and a numerical value and a unit (which might be omitted for
unit-less values). \LaTeX{}-notation symbols other than characters of the Latin alphabet
can easily be included. \textit{Quantities} also support the specification of measurement
uncertainties and estimation errors (cf. Tab.~\ref{tab:Quantities}).
The last item shown in section \lstinline![parameters]! is boolean,
indicating that the measurement was carried out in 4-wire mode.

First analysis results derived from the raw data of solar cell pixel 9
on substrate S419 are listed in the section \lstinline![fingerprints]!. As such data is redundant, but can be very helpful for a quick overview and processing of the recorded data.

The last two sections, \lstinline![*data definitions]! and \lstinline![*data]! differ from the preceeding sections before: the $n^{\rm{th}}$ line of \lstinline![*data definitions]! 
describes the $n^{\rm{th}}$ column of the following \lstinline![*data]! section containing tabular measurement data. The format of the column description is chosen to resemble a typical axis label having a name, a symbol, and a
unit in brackets. In addition, the functional relation of the tabulated
quantities is given by explicitly denoting current $I(V)$ being measured in
dependency on voltage $V$.

\section{A More Complex Example: Documenting Experiment and Analysis Together}
\label{sec:errors}

Applying the basic example of Sec.~\ref{sec:case_study_I} to other
data sets quickly reveals, that for general purposes a more capable
syntax is often needed. For example measurement errors have to be
specified or more than one table may be needed for a comprehensive
description of the data sets. 

\begin{figure}
  \centering
  \psfrag{ldo}{$\ldots$}
  \includegraphics[width=0.8\textwidth,bb=95 73 567 755]{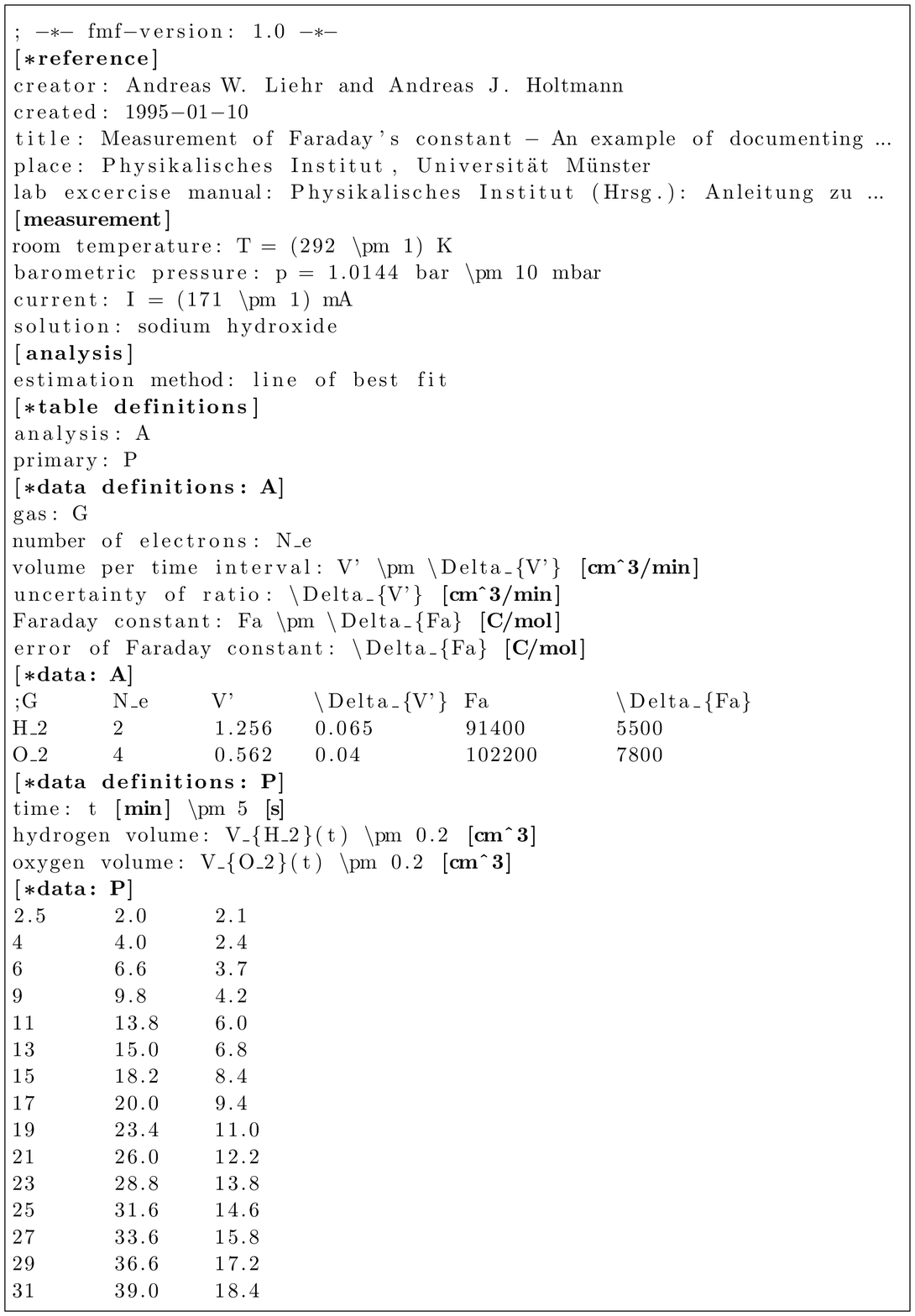}
\caption{\label{fig:FMF-tables}Measurement of Faraday's constant - An
  example of documenting experimental data and their analysis within
  one \fmf-file \cite{Liehr2009_SI20090303b}.
}
\end{figure}

An example of an \fmf-file with two tables is shown in
Fig.~\ref{fig:FMF-tables}. It documents the work of two students in
measuring Faraday's constant in the course of a practical
exercise \cite{Liehr2009_SI20090303b}. The experiment relies on
Faraday's second law and uses a Coulometer for measuring the  
volume fractions of hydrogen and oxygen evolving due to a
constant current $I$ being applied to an aqueous solution of
sodium hydroxide. 
From these time series Faraday's
constant can be computed by converting the volume fractions to normal
conditions (1023mbar and 273K), estimating the evolved
volume per time interval $V^\prime$ from the time series, and
evaluating 
\begin{equation}
  \label{eq:Faraday}
  \mathrm{Fa} = 22.4 \frac{\ell}{\mathrm{mol}}\cdot\frac{I}{N_e V^\prime}
\end{equation}
both for hydrogen and oxygen. Therefore, room temperature and 
barometric pressure at the time of the experiment comprise important metadata for evaluating the measurement. These physical quantities
are specified in section \lstinline![measurement]! of
the \fmf-file shown in Fig.~\ref{fig:FMF-tables} together with their
measurement uncertainties:
\begin{verbatim}
room temperature: T = (292 \pm 1) K
barometric pressure: p = 1.0144 bar \pm 10 mbar
\end{verbatim}
This section also notes the current $I$, which is applied to the sodium
hydroxide solution, and its measurement error. Note that the error
specification is very similar to the way in which a 
scientist would describe the data in a report. 
Other possibilities for specifying errors are listed in
Tab.~\ref{tab:Quantities} of the Appendix.

Because the experiment deals with two different gases, namely
hydrogen and oxygen, which differ in terms of their number of
electrons $N_e$ per reaction, Faraday's constant is individually
retrieved for each time series. Therefore, two tables are needed for
adequately describing the experiment: one table specifying the material parameters and the
result of the data analysis and another table listing the time series of
measured volume fractions. The names of these tables as well as the
associated symbols are defined in section 
\lstinline![*table definitions]! of the \fmf-file
in Fig.~\ref{fig:FMF-tables}. It 
tells that the table named \textit{analysis}, $A$, is followed by the
table \textit{primary}, $P$. Each table then consists of sections
\lstinline![*data definitions: !$X$] and \lstinline![*data: !$X$]
 with $X$ referencing the 
symbol of the table such that each pair can easily be identified.

\begin{figure}
  \centering
  \includegraphics[width=0.618\textwidth]{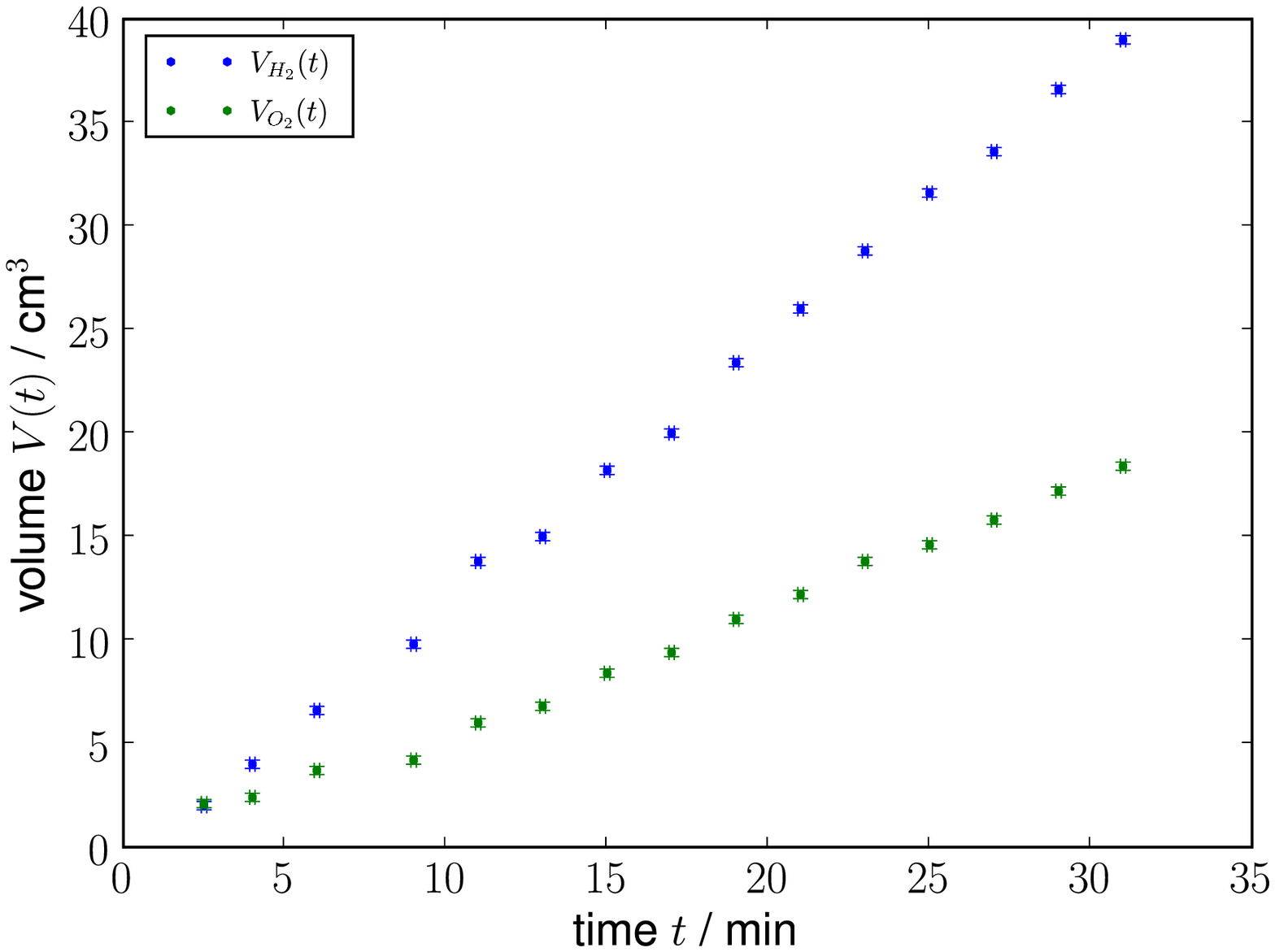}
  \caption{\label{fig:Faraday}Measurement of Faraday's constant.
The diagram visualises the table of data documented in section
\lstinline![*data: P]! of Fig.~\ref{fig:FMF-tables} and uses information from
section \lstinline![*data definitions: P]! to label the graph accordingly \cite{Liehr2009_SI20090303b}.
}
\end{figure}

In this example, two cases of error specifications are needed in the tables: namely
specifying constant measurement errors valid for elements
of a specific column, and assigning special error columns.
The specification of constant measurement errors is shown in section
\lstinline![*data definitions: P]! of the second table in
Fig.~\ref{fig:FMF-tables}:
\begin{verbatim}
time: t [min] \pm 5 [s]
hydrogen volume: V_{H_2}(t) \pm 0.2 [cm^3]
oxygen volume: V_{O_2}(t) \pm 0.2 [cm^3]
\end{verbatim}
In the example, time $t$ is measured in units of minutes with an
accuracy of 5 seconds and volumes $V_{H_2}(t)$ and
$V_{O_2}(t)$ are measured in units of cm$^3$ with an accuracy of
$0.2\mathrm{cm}^3$. With this information at hand, the primary data
of  section \lstinline![*data: P]! can be plotted as shown in
Fig.~\ref{fig:Faraday}.

The specifications of non-constant errors are shown for $V^\prime$ and
Fa in Fig.~\ref{fig:FMF-tables}. The errors  
$\Delta_{V^\prime}$ and $\Delta_{Fa}$, respectively, are 
defined in section \lstinline![*data definitions: A]! and are
explicitly related to the measured quantity as:
\begin{verbatim}
Faraday constant: Fa \pm \Delta_{Fa} [C/mol]
error of Faraday constant: \Delta_{Fa} [C/mol]
\end{verbatim}
These data definitions mean that the column listing the 
Faraday-constant is followed by a column with the corresponding
measurement error. Because this table consists of six columns, the
creator of the \fmf-file decided that the readability would be improved
by starting section \lstinline![*data: A]! with a comment repeating
the symbols defined in section \lstinline![*data definitions: A]!.
The comment is introduced by a leading semicolon.

Alltogether, sections
\lstinline![*data definitions: A]! and \lstinline![*data: A]! of 
Fig.~\ref{fig:FMF-tables} give a simple textual representation of
Tab.~\ref{tab:analysis}, which could be the summary of an experiment.
Section \lstinline![*data definitions: A]! lists the name of the gas, the number $N_e$ of
electrons per reaction, the ratio $V^\prime$ of released gas per time interval and the resulting Faraday constant.
The estimated values of Faraday's constant depend on the number $N_e$ of electrons per reaction.
As can be seen from the last column of Table A in
Fig.~\ref{fig:FMF-tables} the measurement
of Faraday's constant deviates up to 6\% from the precise
value of Fa=96485.3399(24) C/mol~\cite{Faraday}, but has been correctly
determined within the error margins. 

In this case the constant has been determined by means of the line of
best fit. Years later it occures to the students (or maybe even their
successors in the practical exercise) that moving a ruler around on a
piece of paper is perhaps not the best way to analyse the data. Since
the primary data is available in a form easily understood, they decide
to redo the analysis with the more sophisticated means of a
least square fit while taking into account, that
the anode is likely to have an oxide layer, which increases
$V^\prime_{O_2}$ in the beginning of the experiment. With
$V^\prime_{H_2}=1.202 \mathrm{cm}^3/\mathrm{min}\pm 1.0\%$ and 
$V^\prime_{O_2}=0.596\mathrm{cm}^3/\mathrm{min}\pm2.2\%$ Faraday's
constant is now determined as $95600\pm1500$ C/mol for H$_2$ and
$96500\pm2700$ C/mol for O$_2$.  Both these values are far more
accurate compared to the original results. This improvement was
possible, because the original information was preserved in a way that
allowed its interpretation. Furthermore it could be understood,
because the method of analysis was indicated.

While this example might seem a trivial, it is common that data
experimentally gathered by one scientist would be useful to another
one years later. Often the first scientist has become unavailable and
the data can no more be found, let alone understood. This is a 
waste of resources that should be reduced.

\begin{table}
  \centering
  \begin{tabular}{cccc}
  \hline
  \hline
  gas & $N_e$&	$V^\prime\pm\Delta_{V^\prime}$ [cm$^3$/min]&  
  $\mathrm{Fa}\pm\Delta_{Fa}$ [C/mol]\\
  \hline
  H$_2$	&2&	$1.256\pm0.065$ &	$91400\pm	5500$\\
  O$_2$	&4&	$0.562\pm0.04$	& $ 102200\pm7800$\\
  \hline
  \hline
  \end{tabular}
  \caption{Formatted analysis table $A$ of the \fmf file shown in
    Fig.~\ref{fig:FMF-tables}. The table lists the name of the gas, the number $N_e$ of
electrons per reaction, the ratio $V^\prime$ of released gas per time interval and the resulting 
Faraday constant. The ratio $V^\prime$ has
  been determined from table $P$ (Fig.~\ref{fig:FMF-tables}) by
  plotting volume fraction against 
  time for each gas (Fig.~\ref{fig:Faraday}) and estimating the line
  of best fit.}
  \label{tab:analysis}
\end{table}

\section{An advanced example: Searching Scientific Data in terms of units}
\label{sec:search}
How to search for certain physical quantities is explained on the basis of
an example given in Tab.~\ref{Tab:Klassifikator}, where we consider the four energy related quantities
pertaining to different experiments; namely 
work $W= 23 \mathrm{kJ}$, 
energy $E= 10\mathrm{keV}$, calorific
value $H=10\mathrm{kcal}$, and power $P=0.01\mathrm{MW}$. 
A classical full-text search cannot reveal
any correlation between their notation \textit{work}, \textit{energy},
\textit{calorific value}, and \textit{energy}. The same holds for
the units kJ, keV, kcal, and MW. Therefore, a question like
\begin{quote}
  
``Which measurements determine in an energy range between one
  thousand and one billion Joule?''
\end{quote}
cannot be formulated as a full-text query. Instead, the
elements of $\mathbb{M}=\{W,E,H,P\}$ have to be identified which are
energies and also lie in the desired interval
$[1\mathrm{kJ},1\mathrm{MJ}]$: 
\begin{equation}
  \mathbb{M}_{E} = \{m \in \mathbb{M} | \mathrm{dim}\; m = L^2 T
  M^{-2}\} \cap [1\mathrm{kJ},1\mathrm{MJ}].
\end{equation}
Here $L$, $T$, $M$ represent the dimensions length, time and mass, respectively.

The computation of this intersection can be carried out by normalising
the elements of $\mathbb{M}$ to basic SI-units and decomposing each quantity 
into an 8-tuple, the elements of which are its measure and the powers of its
dimensions. In terms of pattern recognition, this 8-tuple is denoted feature
vector. E.g. a physical quantity $q$ is uniquely characterised by 
its feature vector 
$\vec{q}_{F}=(q_0,\ldots,q_7)$ with
\begin{equation}\label{Eq:FeatureVector}
  q = q_0\cdot \mathrm{m}^{q_1}\cdot 
      \mathrm{kg}^{q_2}\cdot 
      \mathrm{s}^{q_3}\cdot 
      \mathrm{A}^{q_4}\cdot 
      \mathrm{K}^{q_5}\cdot 
      \mathrm{mol}^{q_6}\cdot 
      \mathrm{cd}^{q_7}.
\end{equation}
Here $q_0=\{q\}$ is the measure of $q$ and $q_1,\ldots,q_7$ define its
unit:
\begin{equation}\label{Eq:PowerOfUnits}
    [q] = \mathrm{m}^{q_1}\cdot 
      \mathrm{kg}^{q_2}\cdot 
      \mathrm{s}^{q_3}\cdot 
      \mathrm{A}^{q_4}\cdot 
      \mathrm{K}^{q_5}\cdot 
      \mathrm{mol}^{q_6}\cdot 
      \mathrm{cd}^{q_7}.
\end{equation}
As regards the example in Tab.~\ref{Tab:Klassifikator}, all
powers except length, time and mass are zero and only
quantities given in units of
$\mathrm{m}^{2}\mathrm{kg}\,\mathrm{s}^{-2}$
$(q_1,\ldots,q_7)=(2,1,-2,0,0,0,0)$ are energies and therefore are
relevant for determining $10^3\leq q_0\leq10^6$. Consequently, from
the quantities listed in Tab.~\ref{Tab:Klassifikator} only the  
quantities work $W= 23 \mathrm{kJ}$ and calorific value
$H=10\mathrm{kcal}$ pertain to experiments determining energies between 1kJ to 1MJ.

This example illustrates how scientific data sets can be made
searchable on the basis of an adequate documentation, such that the
documentation of data sets directly enables the re-usability of
scientific results.

\begin{table}
  \centering
  \begin{tabular}{lrrrrrrrrrrrrr}
    \hline
    \hline
    &\multicolumn{5}{l}{feature vector $\vec{q}$ (\ref{Eq:FeatureVector})}\\
    \cline{2-9}
    meta-information & $q_0$ & 
    $q_1$& $q_2$ & $q_3$ & $q_4$ & $q_5$& $q_6$& $q_7$\\
    \hline
      work: $W$ = 23 kJ & $23\cdot10^{3}$ & 2 & 1 & -2 & 0 & 0& 0& 0 \\
     energy: $E$ = 10 keV & $1,602\cdot10^{-15}$& 2 & 1 & -2 & 0& 0& 0& 0 \\
     calorific value: $H$ = 10 kcal & $41,9\cdot10^{3}$ & 2 & 1 & -2 & 0& 0& 0& 0\\
     power:  $P$ = 0.01 MW & $10\cdot10^{3}$  & 2 & 1 & -3 & 0& 0& 0& 0  \\
     search interval & $[10^3,10^6]$ & 2 & 1 & -2 & 0 & 0& 0& 0\\
    \hline
    \hline
  \end{tabular}
  \caption{\label{Tab:Klassifikator}Classification of physical
    quantities by means of feature vectors. The feature vector $\vec{q}$
    (\ref{Eq:FeatureVector}) of quantity $q$ is an 8-tuple, 
  which is composed from the measurand $q_0=\{q\}$ in basic SI units
  and its dimension coded as powers of units
  (\ref{Eq:PowerOfUnits}).  Feature vectors with identical elements
  $q_1,\ldots q_7$ correspond to the same physical quantity.}
\end{table}

\section{Discussion}
We have shown how a text file can used as scientific data format enabling storage of
tabular data sets in a consistent and self-descriptive fashion.
The real novelty of the presented data format is its systematic way in which all relevant metadata 
needed to understand the data can seemlessly be included. In language of the
data-information-knowledge-wisdom hierarchy \cite{Ackoff_DIKW} this
means that the data set is upgraded from the data level to the
information level. The promotion to the information level has
significant advantages: 

First, it improves the capability of scientists to communicate scientific data. This
may occur within a working group, with external cooperators, or within the
scientific community in general. Because unhindered communication is one of the most
important preconditions for a successful collaboration, this aspect cannot be
overestimated. Second, it facilitates the long-term integrety of scientific data; e.g if primary data from old project
must be revisited many years after or if data is passed on to comimg generations of scienctists.
At present, it is rather the rule than the exception that the scientist is not able to find all relevant metadata to
understand an old data set. Often such data-erosion is simply due to the meta data residing on a different data storage medium than the
primary data itself. Working with a data format which embodies the relevant metadata avoids this problem altogether.

Using a self-describing format like the one presented here therefore increases the longevity of
primary data and thus may improve the quality of science in general. 
Especially in scientific communities like the geo-sciences or high-energy physics this has proven to be the case.
In these fields, large data sets and the pressure to communicate them effectively has 
led to a standardisation of data formats and a culture of sharing such data.
Due to the complexity of the data generated in those fields more
sophisticated file formats such as HDF5, netCDF or ROOTS \cite{hdf5,netcdf,Root2008_5.21} are in use.

In contrast to these complex data formats, the \fmdf{}
is designed with the needs of so-called 
\textit{Small Science} \cite{onsrud2007} in mind. Research by small working groups and
individuals producing simple tabular data still occupies a central
position in most scientific disciplines. Although the awareness of a
systematic management and sharing of data is already rising, an
appropriate data format for \textit{Small Science} has yet to fully evolve.
One obstacle is that data documentation using the existing extensible markup languages (XML)
like XDF \cite{Shaya2001_XDF} or VOTables \cite{Ochsenbein2004_VOTables} simply add
too much overhead to the content, and are cumbersome to read, edit,
and process with existing scientific software tools.

The approach presented in this paper is simple: Describing simple tabular sets of data with
simple text files in a way which is natural for scientists and
engineers requires a minimum of change in the individual workflow and
habits. In general this means documenting the metadata in a way one
would like to read it in a laboratory journal or in a paper,
e.g. within a diagram or a figure 
caption. The use of plain text files ensures that the scientist can
apply this documentation technique instantaneously with basic
information technology infrastructure. Still, these text files can be
parsed in a very easy way due to their simple structure
\cite{Zimmermann2008_grammar}. 

Because of its simplicity and the self-describing character the \fmdf{} offers many possibilities:
\begin{itemize}
  \item The clear-text documentation of scientific data simplifies its
    re-usage.  
  \item The usage of plain text files makes the data ideal for long term preservation \cite{Rog_LTP}.
  \item The communication of the data does not need a complex infrastructure; text files can be sent by email or even be printed to analogue media.
  \item Because the data is connected to the relevant units, special
    software which is able to process these units during scientific
    data analysis like Pyphant
    \cite{Pyphant,Zimmermann2008_grammar} can be used sucht that processing
    and visualisation of the data can be automated. 
  \item Furthermore, the use of the relevant units enables a semantic
    search within a collection of data sets.
\end{itemize}

A drawback of the \fmdf results from the fact, that the
end-of-line (EOL) character of text files is not uniquely defined for
all operation systems, which causes text files
to be displayed incorrectly after being transfered to a different
type of operation system. However, this problem is generally known and
appropriate tools are available \cite{wiki2009_newline}. 

\section{Conclusion}
The advantage of the suggested file format is its
ease of use and its scientist-friendly syntax, which is in contrast
to the computer-friendly syntax of markup-languages. The purpose of
the \fmdf{} is to document small tabular data sets, mainly
produced in fields generalised as \textit{Small Science}. For these
scientific communities, the use of the \fmdf{} can be the starting point
to a systematic management of scientific data in the form of information, and thus
the starting point for participation in the growing culture of data sharing. 

The authors would like to encourage the reader to engange in the
application of the \fmdf{} and to actively participate in the
improvement of the proposed format.

\clearpage
\newpage

\appendix
\section{The Syntax of the \fmdf}
\label{sec:syntax}
\newcommand{\Sec}[1]{\textbf{#1}}
\newcommand{\namedSec}[1]{\Sec{[#1]}}
\newcommand{\namedKey}[1]{\textit{#1}}

The appendix comprises a more technical description of the syntax
characterising the \fmdf. It is meant as a guide to the format and
shows comprehensive tables of coding examples. Therefore the
appendix intentionally repeats certain parts of the format in order to
minimise browsing for a specific piece of information.

\begin{figure}[hb]
  \centering
  \includegraphics[width=\textwidth,bb=101 554 547 755]{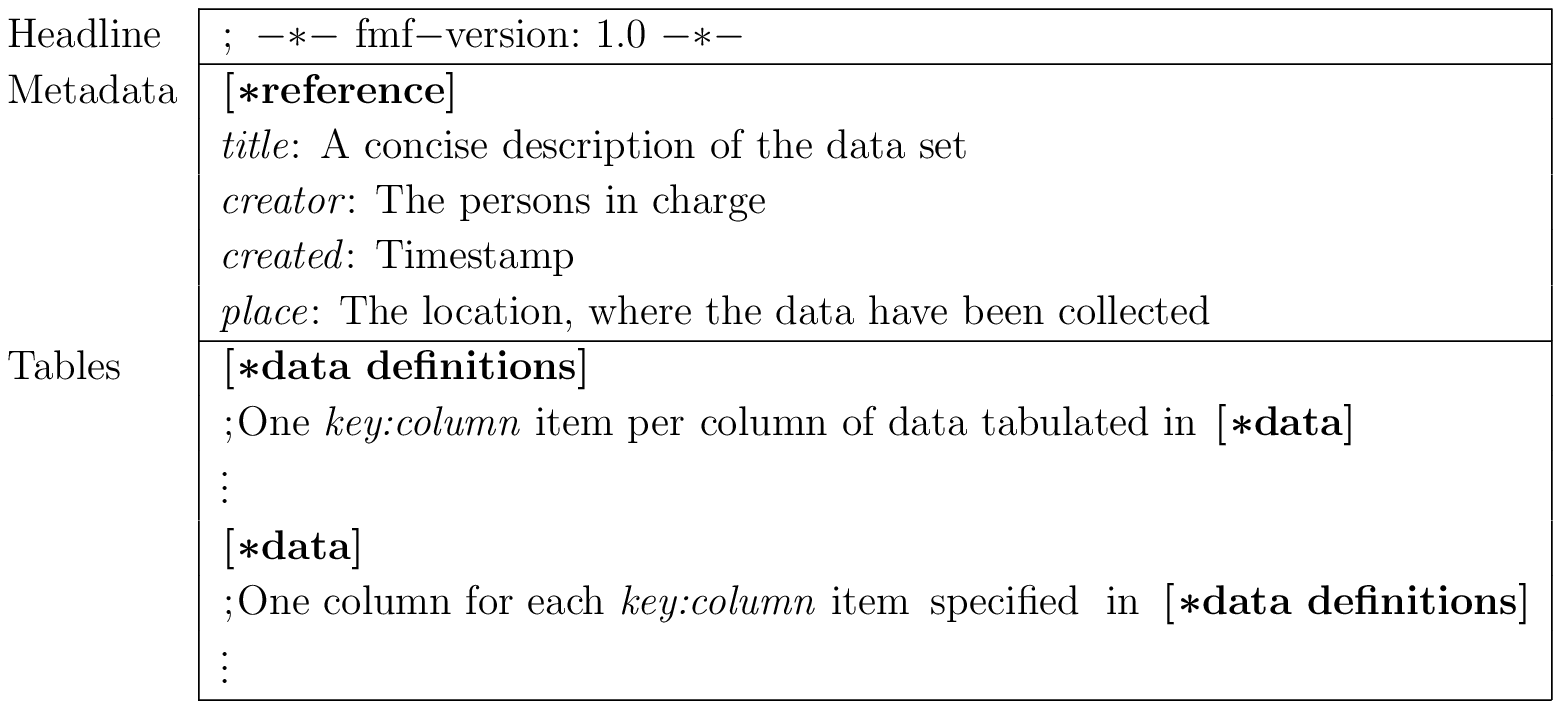}
  \caption{Minimal structure of a \fmdf-file.}
  \label{fig:minimal}
\end{figure}

\begin{figure}[t]
  \centering
  \includegraphics[width=\textwidth,bb=101 258 587 755]{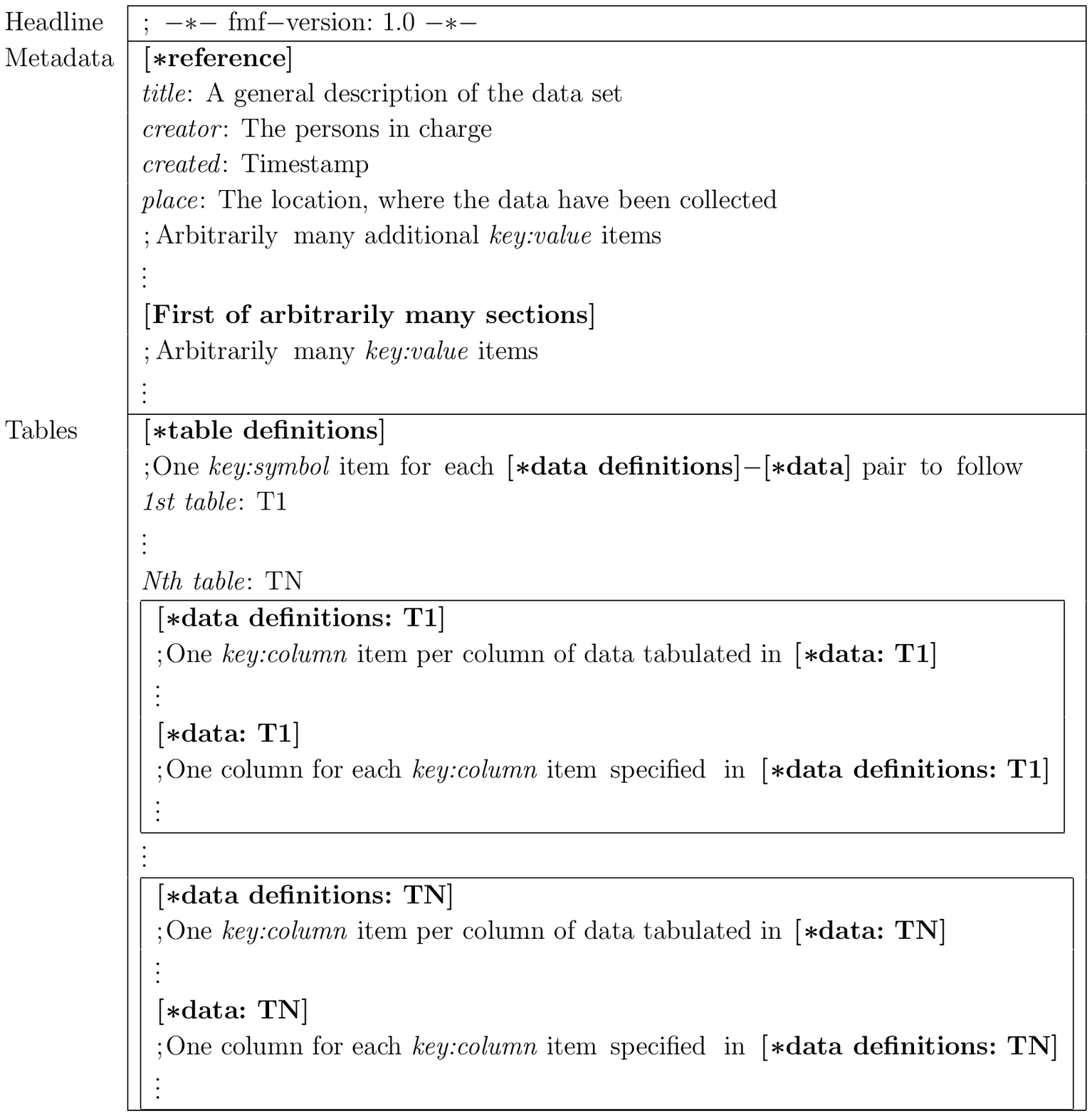}
  \caption{The general structure of an \fmf-file. The headline is used
    to define the file coding and delimiter in the tables, of which
    several are present in the file.}
  \label{fig:general}
\end{figure}

Data files written in the \fmdf always consist of three parts:
\begin{description}
\item[Headline] (\ref{sec:headline}),
\item[Metadata] (\ref{sec:metadata}),
\item[Tables] (\ref{sec:tables}).
\end{description}
The headline is a comment indicating how to interpret the file on a
formal level. Following the headline is the main body of the file,
which is structured in sections. While the file body can contain
arbitrarily many sections with metadata and measurement data, at least
three mandatory sections are needed for a meaningful \fmf-file. These sections
are named \namedSec{*reference}, \namedSec{*data definitions} and
\namedSec{*data}. The \namedSec{*reference} section contains the
metadata necessary for referencing the data set and the
\namedSec{*data definitions} and \namedSec{*data} sections represent a
table of data. This minimal structure is shown in
Fig.~\ref{fig:minimal}, while the general structure is summarized in
Fig.~\ref{fig:general}.

>From a grammatical point of view, the \fmdf consists only of three
different types of lines, defined as follows:
\begin{description}
\item[Comments] are indicated by a leading semicolon (;) or a
  leading sharp (\lstinline!#!). The comment character used for the headline
  (\ref{sec:headline}) has to be used consistently for all other
  comments in the same file. A comment character in a key or value
  is treated as a normal character.
\item[Section headers] are embraced by square brackets $[\;]$
  and have to be unique throughout the file. Section names starting
  with an * are reserved for use in this specification or any future
  version thereof. Any other legal character sequence can be used for
  arbitrary sections. In this version, the following reserved sections
  are put to use:
  \begin{itemize}
  \item \namedSec{*reference},
  \item \namedSec{*table definitions},
  \item \namedSec{*data definitions}, and
  \item \namedSec{*data}.
  \end{itemize}
\item[\namedKey{Key:value} items] are used in all but the
  \namedSec{*data} section. A \namedKey{key} can consist of all
  characters except the colon (:), which is used to separate
  \namedKey{key} and \namedKey{value}. Each \namedKey{key} has to be
  unique within its section. The different types of \textit{values} are
  discussed in \ref{sec:metadata}. In the \namedSec{*reference} and
  all user defined sections, arbitrary value types may be used. The
  \namedSec{*table~definitions} section may only contain
  \namedKey{symbols} as values and the \namedSec{*data~definitions}
  sections only \namedKey{column~specifications}, both for reasons
  that will become clear later on.
\item[Rows of data] are collected in \namedSec{*data} sections. They
  represent classical tabulated data sets. Other column separators than
  tab stop can be specified in the headline (\ref{sec:headline}).
\end{description}

\subsection{Headline}
\label{sec:headline}
The headline is a special comment, which indicates how the content
of the file is to be interpreted. This includes foremost the encoding,
which tells the computer how to translate the bytes of the file into
characters and the separator, which splits the table rows of the
\namedSec{*data} section into the appropriate cells. It also
mandatorily specifies the version of the \fmdf employed in the file.
It uses the Emacs style file syntax \cite{Lemburg2007_UTF8} and thus
looks like
\begin{lstlisting}
; -*- fmf-version: 1.0 -*-
\end{lstlisting}
In addition, \namedKey{coding} (default = utf-8) and
\namedKey{delimiter} (default=tab) can be specified
(Tab.~\ref{tab:headlinevariables}). The \namedKey{key:value} items
have to be separated by a semicolon. Although the semicolon (;) is the
default comment character, comments can alternatively be introduced by
a hash (\#). The comment character used in the headline has
to be used throughout the file.

\subsection{Metadata}
\label{sec:metadata}
Metadata is an essential part of the data file, because it describes
the context from which a data set has been collected. It is
structured by sections, which start with a unique section header
consisting of a section identifier enclosed in square brackets.
Section identifiers starting with an asterisk are reserved for this or
any future version of this specification.
All section headers except the \namedSec{*data}-header are followed by
lines of
\begin{lstlisting}
  key: value
\end{lstlisting}
The key can contain any valid character except a colon, which
separates \namedKey{key} and \namedKey{value}. The value is always a
textual representation of some information. However, in order to allow
for an automated interpretation of the information the \fmdf defines
some conventions for the representation of numerical and boolean
values, quantities and complex strings:
\begin{description}
\item[Boolean values] are given by the words ``true'' or
  ``false''. They can be written in lower case letters, capital letters, or
  with a starting capital letter. A list of boolean values is defined by
  separating the individual values by commas.
\item[Numerical values] are textual representations of integer, real
  or complex scalars. Due to the restrictions of floating point
  arithmetics, the accuracy of real and complex scalars is restricted by
  the number of bits used for encoding the scalar. Optionally, a
  numerical value can be complemented by an uncertainty specified in
  common scientific notation. Furthermore a numerical value can be
  annotated by a symbol in \LaTeX-notation, which is prefixed to the
  number and is related to the latter by an equal sign. A list of
  numerical values is defined by separating the individual values by
  commas. A comprehensive list of possible numerical formats is given in
  Tab.~\ref{tab:numbers}.
\item[Quantities] are measurands, estimations, or control parameters
  of an experiment or simulation. They are characterised by a numerical
  value and a unit. Units are extensively described in
  \ref{sec:units}. A list of quantities is defined by separating the
  individual quantities by commas. A comprehensive list of examples of
  quantities is given in Tab.~\ref{tab:Quantities}.
\item[Timestamps] are ISO formatted date-time strings
  \cite{Kuhn2004_datetime}, for example "2006-04-17 18:55:38+02:00" for
  17th of April 2006 with 18:55:38 local time, which is 2 hours ahead of
  UTC (Tab.~\ref{tab:DateTime}). If the time zone information is omitted,
  the local time zone is assumed. However, in view of international
  cooperations the reference to UTC should always be included. A
  timestamp can also be admended by an uncertainty, which is indicated
  by \lstinline!+-! and a temporal quantity. This is useful for applications of legal
  medicine \cite{Bohnert2008}. A list of
  timestamps is defined by separating the individual timestamps by
  commas.
\item[Strings] are the most flexible type of \namedKey{values} to be
  returned, because a string of characters can map any textual
  information. In particular this applies if the mapping to boolean
  values, numerical values, quantities, or timestamps does not
  match. In order to prevent the interpretation of a textual
  \namedKey{value} in terms of numerical values or quantities the
  information can always be enclosed in quotation marks. However, for
  more complex strings like multi-line strings, lists of strings or
  strings containing quotation marks, some conventions have to be met,
  which are listed in Tab.~\ref{tab:strings}.
\end{description}

\subsection{Tables}
\label{sec:tables}
The \namedSec{*tables} section is a means to include more than one
table in a single \fmf-file. This creates the need to identify
corresponding \namedSec{*data definitions} and \namedSec{*data}
sections. To this end, each table is assigned a name and a symbol,
which in turn is used for identifying the table throughout the
file. This information is found in the \namedSec{*tables} section.
The relevant sections for multi-table files are:
\begin{description}
\item[\namedSec{*table definitions}] This section has one
  \namedKey{key:symbol} item per table. While the \namedKey{key} acts as
  a descriptive name for the table, the \namedKey{symbol} is used to
  relate the \namedSec{*data~definitions} and the \namedSec{*data}
  sections to each other. Therefore these sections reference the table
  $symbol$ within their section header as
  \namedSec{*data~definitions:~symbol} and \namedSec{*data:~symbol}. The
  \namedSec{*table~definitions} section can be skipped, if only one
  table is given within the \fmf-file. In this case, the data definitions
  and the data sections do not reference a symbol and thus are captioned
  by \namedSec{*data definitions} and \namedSec{*data} (see
  Fig.~\ref{fig:minimal}). In general, \LaTeX{}-notation for symbols is
  allowed.
\item[\namedSec{*data definitions}] These sections describe the
  columns of data given in the respective \namedSec{*data} sections by
  means of \namedKey{key:column} items. The $n^\mathrm{th}$ item of a
  \namedSec{*data~definitions} section describes the $n^\mathrm{th}$
  column in the \namedSec{*data} section. A \namedKey{column} value
  specifies a \namedKey{symbol} referencing the tabulated
  quantity. Optionally, it can also define the functional dependency on
  another quantity, a unit and an uncertainty, which is either constant
  or might be tabulated in another column. For the details refer to
  Fig.~\ref{fig:datadefinitionexamples}.
  \begin{figure}[t]
    \centering
    \includegraphics[width=0.618\textwidth,bb=101 438 442 755]{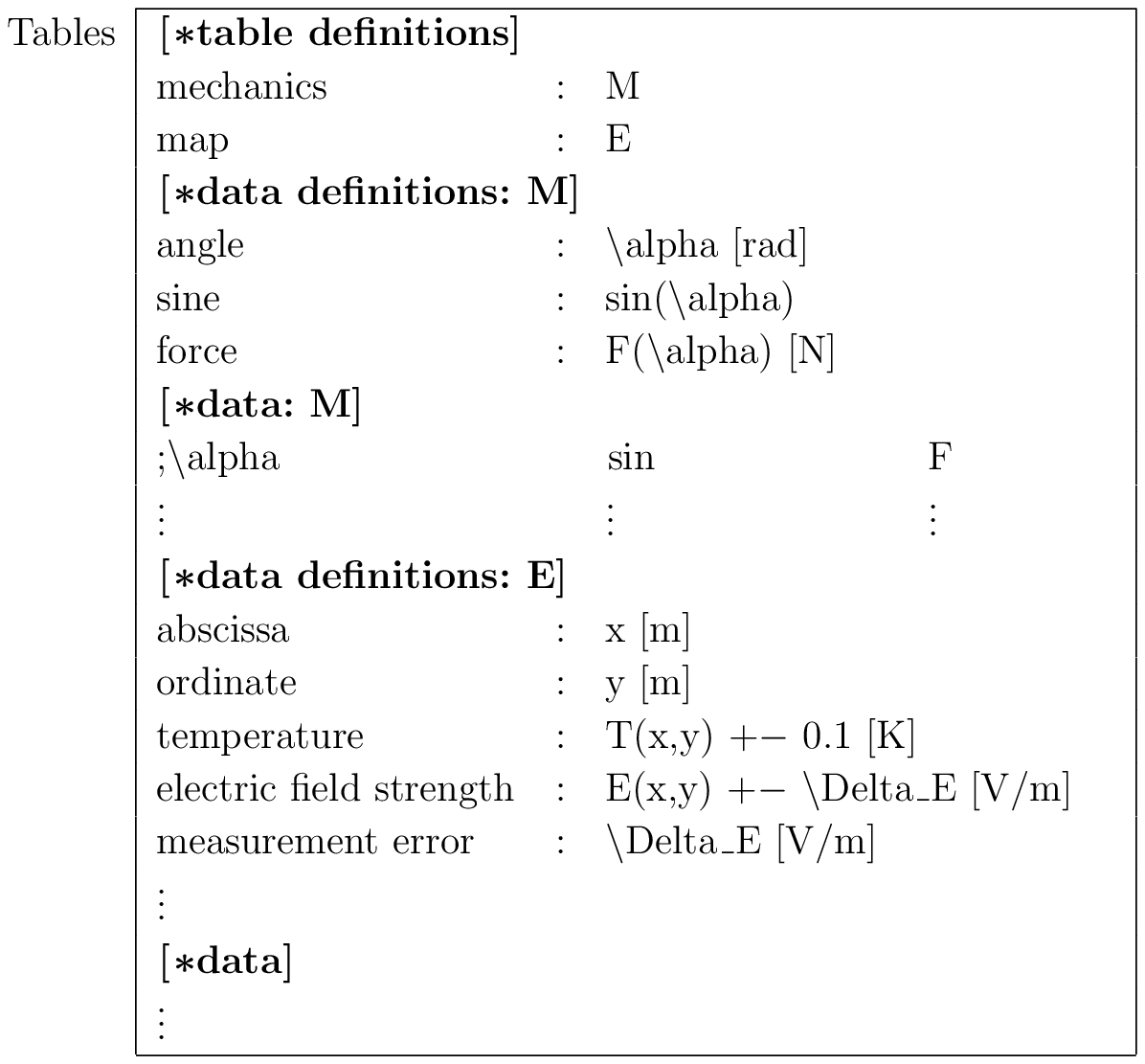}
    \caption{\label{fig:datadefinitionexamples}Structure of the tables
      part comprising two tables.}
  \end{figure}
\item[\namedSec{*data}] These sections are tables of data as shown in
  Figs.~\ref{fig:FMF_solar_example} and \ref{fig:FMF-tables}. The
  columns can contain strings, numerical values, and quantities, whose
  symbols and names are defined in section
  \namedSec{*data~definitions}. The same holds for uncertainties and
  units.  By default, columns are separated with tabs. Other delimiter
  like \lstinline!whitespace!
  can be explicitly defined in the header line of the \fmf-file
  (Table~\ref{tab:headlinevariables}).
\end{description}

\begin{table}[b]
  \centering
  \begin{tabular}[t]{l@{\hspace{1ex}:\hspace{1ex}}ll}
    \hline
    \hline
    Variable & Value & Status\\
    \hline
    fmf-version & 1.0 & Mandatory. Version presented in this paper is 1.0. \\
    coding & utf-8 & Default character encoding \cite{Yergeau2003_UTF8}.\\
    &cp1252 & Example for character encoding with WinLatin1
    code page \cite{Microsoft2005_cp1252}.\\
    delimiter &\lstinline!\t! & Default delimiter is tab.\\
    &\lstinline!whitespace! & Example for column separation by
    whitespaces.\\
    & \lstinline!semicolon! & Example for column separation by
    semicolons (;).\\
    & \lstinline!,! & Example for column separation by commas.\\
    \hline
    \hline
  \end{tabular}
  \caption{Variables defined in the headline. Comprehensive
    information on alternative code pages can be found at
    \cite{Texin2005_codepages}.}
  \label{tab:headlinevariables}
\end{table}
\begin{table}[p]
  \centering
  \begin{tabular}[t]{l@{\hspace{1ex}:\hspace{1ex}}l}
    \hline
    \hline
    Explaining key & Value\\
    \hline
    Integer & 1\\
    Negative integer & -2 \\
    Floating point number & 1.0\\
    Floating point number with leading decimal dot &  .1\\
    Floating point number  with exponential &  1e-10\\
    Another floating point number  with exponential &  -1.1E10 \\
    Complex number& 1+2j\\
    Another complex number&  1.1+2J \\
    Complex number with zero real part & 2J \\
    Complex number with zero imaginary part& 1+0J\\
    List of floats&     1.0, .1, 1e-10, -1.1E10\\
    Parameter & P = 42.0 \\
    Parameter with uncertainty & Q = 42.1 \lstinline!+-! 0.2 \\
    Parameter with relative uncertainty  & Q' = 42.1 \lstinline!+-! 0.48\% \\
    \hline
    \hline
  \end{tabular}
  \caption{\label{tab:numbers}Examples for textual representations of
    scalars. A value is interpreted as integer if the respective string
    contains only digits and an optional leading sign. A string is
    interpreted as floating point number if it contains a decimal dot or
    an exponent indicated by an embedded 'e' or 'E'. Complex numbers are
    coded as a sum of real and imaginary parts in integer or floating point
    notation. The imaginary part is indicated by a trailing 'j' or
    'J'. Lists of numbers are built from comma separated numbers. Special
    values like \lstinline!NaN! (not a number) or \lstinline!+INF! and
    \lstinline!-INF! for $\pm\infty$ are also allowed (IEEE
    754)~\cite{ieee-754}. Optionally
    numerical values can be complemented by uncertainties and a symbol in
    \LaTeX{}-notation. Note that the uncertainty sign can also be
    given by $\backslash$pm. }
\end{table}
\begin{table}[p]
  \centering
  \begin{tabular}[t]{l@{\hspace{1ex}:\hspace{1ex}}l}
    \hline
    \hline
    Explaining key & Quantity\\
    \hline
    Physical quantity & 2.0 ohm\\
    & 2.0 kg*m**2/A**2/s**3\\
    & 2.0 kg*m$\hat{\;}$2/A$\hat{\;}$2/s$\hat{\;}$3\\
    & 2.0 kg*m$\hat{\;}$2*A$\hat{\;}$-2*s$\hat{\;}$-3\\
    Physical quantity with uncertainty &  2.0 ohm \lstinline!+-!
    0.02 ohm \\
    &  2.0 ohm  \lstinline!+-! 20 mohm \\
    &  (2.0  \lstinline!+-! 0.02) ohm\\
    & (2.0  \lstinline!+-! 1 \%) ohm\\
    & (1.0  \lstinline!+-! 0.01) 2.0 ohm\\
    & (1.0  \lstinline!+-! 1\%) 2.0 ohm\\
    Monetary quantity &  19.99 EUR/m**2\\
    List of quantities & 2.0 ohm, 2.0 ohm \lstinline!+-! 0.02 ohm,
    19.99 EUR/m**2\\
    Resistance & R = 2.0 ohm\\
    Temperature & $\backslash$theta = 32.0 K\\
    Measured resistance &  R = 2.0 ohm \lstinline!+-! 0.02 ohm \\
    \hline
    \hline
  \end{tabular}
  \caption{\label{tab:Quantities}Examples for textual representations
    of \textit{quantities}. They are specified by a numerical value
    (Tab.~\ref{tab:numbers}) and a unit (\ref{sec:units}). Optionally,
    quantities can be complemented by uncertainties and a symbol in
    \LaTeX{}-notation. Note that the uncertainty sign can also be given
    by $\backslash$pm.}
\end{table}
\begin{table}[p]
  \centering
  \begin{tabular}[t]{l@{\hspace{1ex}:\hspace{1ex}}l}
    \hline
    \hline
    Explaining key & Value\\
    \hline
    date &  2008-12-16\\
    week date& 2008-W47-1\\
    date-time& 2008-12-16T16:51\\
    another date-time& 2008-12-16 16:51\\
    date-time with seconds& 2008-12-16T16:51:05\\
    date-time UTC& 2008-12-16T16:51Z\\
    date-time+2h& 2006-04-23 14:25:51+02:00\\    
    date-time with uncertainty& 2008-12-16 16:30\lstinline!+-!2 h\\
    list of dates& 2008-11-17,2008-1-3,2006-·2-17,2008-W47-1\\
    \hline
    \hline
  \end{tabular}
  \caption{\label{tab:DateTime}Examples for ISO formatted date-time
    strings \cite{Kuhn2004_datetime}.}
\end{table}
\begin{table}[p]
  \centering
  \begin{tabular}[t]{l@{\hspace{1ex}:\hspace{1ex}}l}
    \hline
    \hline
    Explaining key & Value\\
    \hline
    Text & Demonstrating the flexibility of the \fmdf\\
    Comma separated list & Freiburger Materialforschungszentrum,
    University of Freiburg\\
    Quoted text & "Freiburger Materialforschungszentrum, University of
    Freiburg"\\
    Single quote & 'Freiburger Materialforschungszentrum, University
    of Freiburg'\\
    Inside quotation& Arthur C. Clarke's "The Sentinel"\\
    Multi-line & \lstinline!'''!A multi-line value, that spans more
    than one line: \\
    \multicolumn{2}{r}{The line breaks are included in the
      value.\lstinline!'''!}\\
    Another multi-line & """A multi-line value, that spans more than one
    line:\\
    \multicolumn{2}{r}{line breaks are included in the value."""}\\
    Enclosed quotation marks & """ "Don't visualise data, document it!"
    """\\
    \hline
    \hline
  \end{tabular}
  \caption{\label{tab:strings}Examples for textual representations of
    information, which are mapped to strings of characters. Text values
    can be quoted by single quote, a single forward apostrophe (') and by
    double quotation marks (") in order to prevent the interpretation of
    the text value by the parser. Triple quotes are used for multi-line
    text values or in cases for which the text value starts and ends with
    quotation marks.}
\end{table}
\clearpage
\section{Units}
\label{sec:units}
Units are defined on the basis of the SI units Metre (m), Kilogram (kg),
Second (s), Ampere (A), Kelvin (K), Mol (mol), Candela (cd), and the
derived units Newton (N), Pascal (Pa), Joule (J), Watt (W), Coulomb
(C), Volt (V), Farad (F), Ohm (ohm), Siemens (S), Weber (Wb), Tesla
(T), Henry (H), Lumen (lm), Lux (lx), Becquerel (Bq), Gray (Gy),
Sievert (Sv), Radiant (rad), and Steradiant (Sr). Moreover,
monetary values can be defined on the basis of the Euro (EUR) exchange
rates as published by the European Central Bank \cite{ECB}.  The order
of magnitude for all units can be specified by metric prefixes
(Tab.~\ref{tab:prefixes}). Constants and additional non-SI units are
listed as follows:
\begin{description}
\item[Tab.~\ref{tab:constants}] Mathematical and physical constants.
\item[Tab.~\ref{tab:time}] Time units.
\item[Tab.~\ref{tab:length}] Length and area units.
\item[Tab.~\ref{tab:volume}] Volume units.
\item[Tab.~\ref{tab:mass}] Mass and force units.
\item[Tab.~\ref{tab:energy}] Energy and power units.
\item[Tab.~\ref{tab:pressure}] Pressure units.
\item[Tab.~\ref{tab:degrees}] Geometrical and thermo-dynamical degrees.
\end{description}
Note that the abbreviation ``a.u.'' is used for \textit{arbitrary
units} and not for atomic units. This is due to the fact, that atomic
units form a system of units in which several physical constants are
defined as unity \cite{ShullHall1959_AtomicUnits}. E.g. for Hartree
atomic units the mass and charge of the electron, the Bohr radius, the
absolute value of the electric potential energy of the Hydrogen atom
in its ground state, Planck's constant and the permittivity of vacuum
are unity by definition, which of course collides with the
searchability of scientific data discussed in Sec.~\ref{sec:search}.
\begin{table}[t]
  \centering
  \begin{tabular}[t]{lll}
    \hline
    Symbol & Prefix & Order of magnitude\\
    \hline
    Y & yotta- & $10^{24}$\\
    Z & zetta- & $10^{21}$\\
    E & exa-   & $10^{18}$\\
    P & peta-  & $10^{15}$\\
    T & tera-  & $10^{12}$\\
    G & giga-  & $10^{9}$\\
    M & mega-  & $10^{6}$\\
    k & kilo-  & $10^{3}$\\
    da& hecto- & $10^{2}$\\
    d & deci-  & $10^{-1}$\\
    c & centi- & $10^{-2}$\\
    m & milli- & $10^{-3}$\\
    mu& micro- & $10^{-6}$\\
    n & nano-  & $10^{-9}$\\
    p & pico-  & $10^{-12}$\\
    f & femto- & $10^{-15}$\\
    a & atto-  & $10^{-18}$\\
    z & zepto- & $10^{-21}$\\
    y & yocto- & $10^{-24}$\\
    \hline
    \hline
  \end{tabular}
  \caption{Prefixes that can be used for base and derived SI units.}
  \label{tab:prefixes}
\end{table}
\begin{table}[p]
  \centering
  \begin{tabular}[t]{lll}
    \hline
    \hline
    Symbol & Value & Description\\
    \hline
    pi & 3.1415926535897931 & Area of unit circle\\
    c  & 299792458.*m/s     & Speed of Light\\
    mu0& 4.e-7*pi*N/A**2    & Permeability of vacuum\\
    eps0&1/mu0/c**2         & Permittivity of vacuum\\
    Grav&6.67259e-11*m**3/kg/s**2 & Gravitational constant\\
    hplanck&6.6260755e-34*J*s& Planck constant\\
    hbar& hplanck/(2*pi)     & Planck constant / 2pi\\
    e   & 1.60217733e-19*C   & Elementary charge\\
    me  & 9.1093897e-31*kg   & Electron mass\\
    mp  & 1.6726231e-27*kg   & Proton mass\\
    Nav & 6.0221367e23/mol   & Avogadro number\\
    k   & 1.380658e-23*J/K   & Boltzmann constant\\
    \hline
    \hline
  \end{tabular}
  \caption{Mathematical and physical constants.}
  \label{tab:constants}
\end{table}
\begin{table}[p]
  \centering
  \begin{tabular}[t]{lll}
    \hline
    \hline
    Symbol & Value & Description\\
    \hline
    min & 60*s & Minute\\
    h   & 60*min & Hour\\
    d   & 24*h & Day\\
    wk  & 7*d&Week\\
    yr  & 365.25*d & Year\\
    \hline
    \hline
  \end{tabular}
  \caption{Time units.}
  \label{tab:time}
\end{table}
\begin{table}[p]
  \centering
  \begin{tabular}[t]{lll}
    \hline
    \hline
    Symbol & Value & Description\\
    \hline
    AU  & 149597870691m & Astronomical unit\\
    Ang & 1.e-10*m & Angstrom\\
    Bohr&4*pi*eps0*hbar**2/me/e**2& Bohr radius\\
    ft  & 12*inch & Foot\\
    inch& 2.54*cm & Inch\\
    lyr & c*yr & Light year\\
    mi  & 5280.*ft & (British) mile\\
    nmi & 1852.*m & Nautical mile\\
    pc  & 3.08567758128e16m & Parsec\\
    yd  & 3*ft    & Yard\\
    acres & mi**2/640 & Acre\\
    b   & 1.e-28*m**2 & Barn \\
    ha  & 10000*m**2 & Hectare \\
    \hline
    \hline
  \end{tabular}
  \caption{Length and area units.}
  \label{tab:length}
\end{table}
\begin{table}[p]
  \centering
  \begin{tabular}[t]{lll}
    \hline
    \hline
    Symbol & Value & Description\\
    \hline
    l & dm**3 & Litre\\
    dl & 0.1*l & Decilitre\\
    cl & 0.01*l& Centilitre\\
    ml & 0.001*l&Millilitre\\
    tsp & 4.92892159375*ml & Teaspoon\\
    tbsp& 3*tsp & Tablespoon\\
    floz& 2*tbsp & Fluid ounce\\
    cup & 8*floz & Cup\\
    pt  & 16*floz& Pint\\
    qt  & 2*pt   & Quart\\
    galUS & 4*qt & US gallon\\
    galUK & 4.54609*l & British gallon\\
    \hline
    \hline
  \end{tabular}
  \caption{Volume units.}
  \label{tab:volume}
\end{table}
\begin{table}[p]
  \centering
  \begin{tabular}[t]{lll}
    \hline
    \hline
    Symbol & Value & Description\\
    \hline
    amu & 1.6605402e-27*kg & Atomic mass units\\
    oz  & 28.349523125*g   & Ounce\\
    lb  & 16*oz            & Pound\\
    ton & 2000*lb          & Ton \\
    dyn & 1.e-5*N& Dyne (cgs unit) \\
    \hline
    \hline
  \end{tabular}
  \caption{Mass and force units.}
  \label{tab:mass}
\end{table}
\begin{table}[p]
  \centering
  \begin{tabular}[t]{lll}
    \hline
    \hline
    Symbol & Value & Description\\
    \hline
    erg & 1.e-7*J & Erg (cgs unit) \\
    eV  & e*V & Electron volt\\
    Hartree & me*e**4/16/pi**2/eps0**2/hbar**2 & Hartree \\
    invcm & hplanck*c/cm & Wave-numbers/inverse cm\\
    Ken & k*K & Kelvin as energy unit\\
    cal & 4.184*J & Thermo-chemical calorie\\
    kcal& 1000*cal& Thermo-chemical kilo-calorie\\
    cali& 4.1868*J &International calorie\\
    kcali& 1000*cali & International kilo-calorie\\
    Btu & 1055.05585262*J & British thermal unit\\
    hp & 745.7*W & Horsepower\\
    \hline
    \hline
  \end{tabular}
  \caption{Energy and power units.}
  \label{tab:energy}
\end{table}
\begin{table}[p]
  \centering
  \begin{tabular}[t]{lll}
    \hline
    \hline
    Symbol & Value & Description\\
    \hline
    bar & 1.e5*Pa & Bar (cgs unit)\\
    dbar & 1.e4*Pa & Decibar (cgs unit)\\
    mbar & 1.e2*Pa & Millibar (cgs unit)\\
    atm & 101325.*Pa & Standard atmosphere\\
    torr& atm/760    & Torr = mm of mercury\\
    psi & 6894.75729317*Pa & Pounds per square inch\\
    \hline
    \hline
  \end{tabular}
  \caption{Pressure units.}
  \label{tab:pressure}
\end{table}
\begin{table}[p]
  \centering
  \begin{tabular}[t]{lll}
    \hline
    \hline
    Symbol & Value & Description\\
    \hline
    deg & pi*rad/180 & Degrees \\
    degR & (5./9.)*[K] & Degrees Rankine\\
    degC & [K]-273.15 & Degrees Celsius\\
    degF & 5./9.*[K]-459.67 & Degrees Fahrenheit\\
    \hline
    \hline
  \end{tabular}
  \caption{Geometrical and thermo-dynamical degrees.}
  \label{tab:degrees}
\end{table}

\clearpage
\section*{Acknowledgements}
The authors would like to thank H. H. Winter, M. Walter, and K. Kaiminsky for fruitful discussions
on the topic. A. W. Liehr greatefully acknowledges the Apple Research
\& Technology Support (ARTS).

\section*{References}

\end{document}